\documentclass[conference]{IEEEtran}
\IEEEoverridecommandlockouts
\usepackage{cite}
\usepackage{amsmath,amssymb,amsfonts}
\usepackage{algorithmic}
\usepackage{graphicx}
\usepackage{textcomp}
\usepackage{xcolor}
\usepackage{float}
\usepackage{balance}
\usepackage{tabularx}
\usepackage{booktabs}
\def\BibTeX{{\rm B\kern-.05em{\sc i\kern-.025em b}\kern-.08em
    T\kern-.1667em\lower.7ex\hbox{E}\kern-.125emX}}

\begin{document}

\title{Towards Human-AI Synergy in Requirements Engineering: A Framework and Preliminary Study}
\author{%
  \IEEEauthorblockN{%
    Mateen Ahmed Abbasi\IEEEauthorrefmark{1}, 
    Petri Ihantola\IEEEauthorrefmark{1}, 
    Tommi Mikkonen\IEEEauthorrefmark{1}, 
    Niko Mäkitalo\IEEEauthorrefmark{1}%
  }
  \\
  \IEEEauthorblockA{%
    \IEEEauthorrefmark{1}%
    \textit{Faculty of Information Technology}, 
    \textit{University of Jyväskylä}, 
    Jyväskylä, Finland\\
    Email: mateen.a.abbasi@jyu.fi; petri.j.ihantola@jyu.fi; tommi.j.mikkonen@jyu.fi; niko.k.makitalo@jyu.fi
  }
}

\maketitle

\begin{abstract}
 The future of Requirements Engineering (RE) is increasingly driven by artificial intelligence (AI), reshaping how we elicit, analyze, and validate requirements. Traditional RE is based on labor-intensive manual processes prone to errors and complexity. AI-powered approaches, specifically large language models (LLMs), natural language processing (NLP), and generative AI, offer transformative solutions and reduce inefficiencies. However, the use of AI in RE also brings challenges like algorithmic bias, lack of explainability, and ethical concerns related to automation. To address these issues, this study introduces the Human-AI RE Synergy Model (HARE-SM), 
a conceptual framework that integrates AI-driven analysis with human oversight to improve requirements elicitation, analysis, and validation. The model emphasizes ethical AI use through transparency, explainability, and bias mitigation. We outline a multi-phase research methodology focused on preparing RE datasets, fine-tuning AI models, and designing collaborative human-AI workflows. 
This preliminary study presents the conceptual framework and early-stage prototype implementation, establishing a research agenda and practical design direction for applying intelligent data science techniques to semi-structured and unstructured RE data in collaborative environments.
\end{abstract}


\section{Introduction}
Requirements engineering (RE) is the cornerstone of the software development lifecycle, dedicated to a process of understanding, analyzing, and managing the functional and non-functional requirements of a system\cite{flores2009requirements,ronanki2023re}. It's the bridge between stakeholder needs and a successful, well-designed software solution. Traditionally, RE uses manual processes, making it labor-intensive and prone to errors \cite{shahbeklu2024requirement,nasim2023requirement}, leading to issues such as:

\begin{itemize}
    \item \textbf{Stakeholder miscommunication} often leads to ambiguity  in requirement gathering\cite{connor2009bridging,de2012elicitation,dalpiaz2020requirements,bano2015addressing}
    \item \textbf{Information overload} due to large volumes of unstructured requirements
    \item \textbf{Inconsistencies and bias} in requirement prioritization and validation\cite{felfernig2021ai,hudaib2018requirements}
\end{itemize}

Recent advances in Data Science and AI, particularly Large Language Models (LLMs) and NLP, have shown the potential to support and automate RE tasks. LLMs such as GPT can extract structured requirements from unstructured data\cite{arora2024advancing}, while ML models can help to classify, prioritize, and validate requirements. AI is revolutionizing RE practices through the automation of complex tasks such as elicitation, analysis, and validation of requirements\cite{dalpiaz2020requirements,cheng2024generative}. However, the integration of AI into RE is not without challenges\cite{arora2024advancing,cheng2024generative}. Automation bias, lack of domain-specific understanding, and ethical issues related to transparency and accountability hinder the adoption of AI-driven RE tools\cite{arora2024advancing}. Despite theoretical advancements, the practical implementation of AI-enhanced tools in RE has not been fully explored, especially in human-AI collaborative environment.\cite{maalej2023tailoring,dalpiaz2020requirements}.


In this paper, we introduce a Human-AI Requirements Engineering Synergy Model (HARE-SM), which is designed to integrate AI-based automation and human judgment while ensuring trust, transparency, and explainability. The vision of the model is to enhance RE efficiency and reliability through the use of AI in requirements elicitation, analysis, and validation, while humans continue to oversee major decision-making.

To achieve this goal, we propose a research plan that involves: 
\begin{itemize}
    \item Operationalizing and iteratively refining HARE-SM through prototype implementation and stakeholder feedback.
    \item Implementing AI-assisted RE techniques (e.g. LLM-based requirement extraction, transformer-based classification and prioritization, NLP conflict detection, generative drafting, and performance benchmarking)
    \item Conducting stakeholder-driven evaluations to assess trust, usability, and fairness. 
    \item Establishing feedback loops to iteratively improve the role of AI in the process.
\end{itemize}


This paper is organized as follows. In Section~\ref{sec:related work}, we outline related research and identifies key literature gaps. Section~\ref{sec:Methodology} presents our research roadmap, including core research questions~\ref{subsec:Rq} and the four‐phase methodology~\ref{subsec:Methodology}. In Section~\ref{sec:HARE}, we introduce the Human-AI RE Synergy Model (HARE-SM). Section~\ref{sec:approach} describes our HARE-SM prototype, detailing its features, system architecture, and model‐support implementation. Section~\ref{sec:Contributions} outlines planned next steps, and Section~\ref{sec:Conclusion} concludes with a summary of contributions and directions for future work.

\section{Motivation and Related Work}
\label{sec:related work}

RE has traditionally been a manual and iterative process, relying on human expertise to elicit, analyze, and validate requirements. The integration of AI into RE presents a unique opportunity to enhance automation and decision support, driving more efficient project outcomes\cite{kaur2020review}. Though AI-based tools, specifically Large Language Models (LLMs) and Natural Language Processing (NLP), are identified to automate RE tasks, there are still challenges that remain unaddressed.

\subsection{Requirements Engineering (RE) and AI Integration}

Traditional requirements engineering methods cannot perfectly manage large amounts of unstructured requirements data, However, these data can be effectively processed by AI algorithms\cite{shahbeklu2024requirement}. Recent advancements in large language models (LLMs) and AI-based algorithms have shown the feasibility of automating the core RE processes.  Arora et al. \cite{arora2024advancing} demonstrated requirement elicitation is facilitated by LLMs as they act as conversational agents that process domain-specific knowledge and generate structured requirements from unstructured data. Furthermore, the use of LLMs can enhance the precision of requirements elicitation and validation, as illustrated in their ActApp case study. 

Luitel et al. \cite{luitel2023} demonstrated the use of pre-trained language models, i.e., BERT, for enhancing the external completeness of requirements. By predicting missing terminology through masked language modeling, their approach facilitates the identification of potentially missing concepts in existing requirements and thereby improves specification quality.

Regardless of the benefits of AI integration into RE, a structured framework is essential which promotes human-AI collaboration. The effectiveness of collaboration depends on understanding the distinct roles of human experts and AI tools. Fragiadakis et al. \cite{fragiadakis2024} examined the establishment of guidelines for interaction accountability can reduce the risks associated with overreliance on AI. They suggest that a human-centric framework needs to be transparent and make it possible for all stakeholders to understand how AI makes decisions and be able to intervene when needed\cite{fragiadakis2024,nazarenko2024}. Bias mitigation in AI systems has also become a pertinent topic when it comes to RE. Ma et al.\cite{ma2023fairness} investigate predictive bias in LLMs during in-context learning, highlighting how prompt variations influence model outputs. The implications of bias in LLM-generated text raise important ethical considerations when such models are applied in RE tasks. 

Fairness-aware algorithms are key to integrating AI in RE, as they detect and mitigate bias in data and model behavior\cite{liu2024faim}. Techniques like reweighing or sampling\cite{valentim2019assessing}, adversarial debiasing\cite{le2020adversarial}, and post-processing adjustments such as equalized odds can help ensure fairness across stakeholders and can build trust in AI-assisted RE.

Despite these advancements, practical adoption of AI-driven RE tools remains limited. A lot of studies concentrate on theoretical solutions, meanwhile, their application in practice is crucial, especially in fast-paced software engineering environments, where AI and humans work closely \cite{laplante2022}.

\subsection{Ethical Considerations \& Bias}

The integration of AI in RE raises ethical concerns, particularly regarding bias and transparency in decision-making\cite{gardner2022ethical}. These concerns can significantly impact the quality and acceptability of software systems. Boch et al. \cite{boch2022} argue that AI systems should be held accountable to ethical norms governing bias and transparency. They suggest that explainability be part of the framework to alleviate the concern of the "black box" related to many AI models, emphasizing that AI-driven decisions will be made known to and biases identified and managed by humans. Arrieta et al. \cite{arrieta2020} share this view, explaining that explainable AI (XAI) systems enhance accountability by providing transparency, enabling stakeholders to understand AI-driven decision-making in RE processes. Ma et al.\cite{ma2023fairness} underscore the potential ethical risks of biased outputs in AI-assisted systems; their findings caution against over-reliance on biased prompts can influence the fairness and quality of AI-generated outputs. To maintain the stakeholder's trust it is necessary to ensure that AI tools fulfill the ethical standards such as GDPR and IEEE guidelines \cite{gdpr2016}. This study will address these issues by showing the existing codes of ethics and measures that can be commenced to minimize harm from the technologies in RE.


\subsection{Need for Human-AI Collaboration in RE}

 Recent studies highlight the importance of human-AI collaboration, emphasizing that while AI can automate tedious and repetitive tasks, human expertise remains essential for validation, navigating complex decision-making, and addressing ethical concerns that AI may overlook\cite{Song_Zhu_Luo_2024,subramonyam2022}. AI technologies allow active participation between different stakeholders and process feedback efficiently, Saha et al. \cite{saha2023human} points out that AI -driven platforms allow stakeholders to state their requirements and other preferences more conveniently, which makes the requirement collection process more comprehensive. These insights motivate the development of structured collaboration frameworks that deliberately balance AI assistance with human judgment throughout the RE lifecycle.
 
 A group of software engineers who have integrated ChatGPT into software development workflows reported that AI accelerates coding and routine analyzes, but human guidance is essential to handle domain-specific nuances and security considerations\cite {hamza2024}. Another perspective comes from an expert panel at RE’23, noting that although LLMs and generative AI are “inescapable” for routine RE tasks, complex and safety-critical systems “still require skilled engineers” in the loop\cite{borg2024requirements}. In other words, AI can automate or assist certain requirements activities, but maintaining human involvement and trust is crucial for high-stakes decisions. With this observation, Mittelstadt et al.\cite{mittelstadt2016} caution against "automation bias in decision-making" which is the consequence of over-reliance on AI, where human input is undervalued. To get the best results in RE, AI and human judgment must complement each other, balancing automation and human intuition.
 

\section{Research Roadmap}
\label{sec:Methodology}

To investigate and enhance human-AI collaboration in RE, we have structured our long-term research around three core research questions and a four‐phase research roadmap. Each phase has a specific research objective and methodology, and contributes to imploring human--AI collaboration in requirements engineering. 
In this paper, we build on our earlier research (Phase I) and report research related to Phase II.

\subsection{Research Questions}
\label{subsec:Rq}

In the proposed research, we aim to address the following research questions:
\begin{itemize}
    \item \textbf{RQ1:} How can AI tools enhance requirements elicitation, analysis, and validation in collaborative human-AI environments?
    \item \textbf{RQ2:} What challenges arise in implementing AI-driven RE tools, and how can these challenges be addressed?
    \item \textbf{RQ3:} What ethical considerations should govern the use of AI in RE to ensure fairness, transparency, and accountability?
\end{itemize}

\subsection{Methodology}
\label{subsec:Methodology}

\begin{figure*}[htbp]  
\centering
\includegraphics[width=\textwidth]{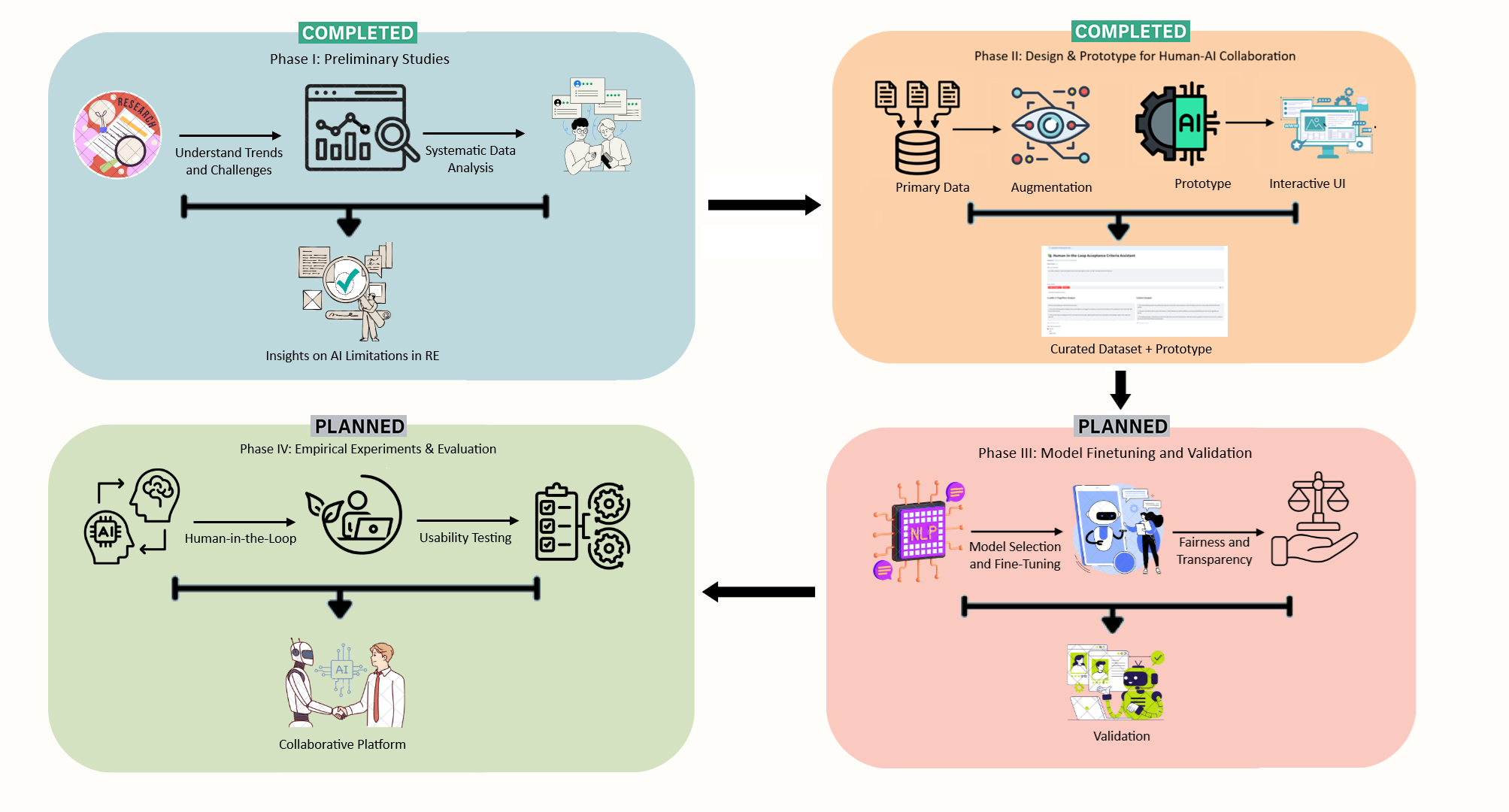} 
\caption{Research phases introduced. 
}
\label{fig:methodology}
\end{figure*}

Our research follows a systematic four‐phase methodology that balances technical feasibility with human‐centric considerations in AI‐supported requirements engineering. This structured process addresses the key challenges identified earlier and lays the groundwork for the conceptual model introduced in Section~\ref{sec:HARE}.
Fig.~\ref{fig:methodology} provides an overview of the research methodology.

Model development begins with conceptual modeling, drawing insights from existing research on AI-enabled RE and human-AI collaboration models. The initial model is established from identified challenges, responsible AI principles, and stakeholders' needs.
To ensure practical validity and relevance, we will engage requirements engineers, AI practitioners, and domain experts. Additionally, key considerations of explainability, trust, and ethical concerns will be assessed at this point to ensure responsible AI integration.

In the following, we introduce the four phases of the research together with their objective, method, and expected research outcomes.

\subsection*{Phase I: Preliminary Studies}

This phase serves as the foundation for developing our framework by analyzing existing AI-driven RE approaches, identifying challenges, and understanding stakeholder expectations. By synthesizing insights from prior studies, we ensure that our model aligns with best practices and addresses the gaps in AI-assisted RE.
\begin{itemize}
    \item \textbf{Objective:} Conduct comprehensive background research to elaborate on challenges and gaps and establish knowledge for the proposed framework.
    \item \textbf{Method:} literature review of peer-reviewed articles, conference papers, and industry reports about AI applications in RE methodologies, tools, and challenges.
    \item \textbf{Expected Outcome:} A synthesized overview of the primary challenges, ethical issues, and limitations of available AI solutions for RE, which will then serve as a foundation for the design of improved frameworks.
    \item \textbf{RQ Mapping:} This phase addresses RQ2 by identifying challenges in AI-driven RE and analyzing prior efforts to mitigate them.
\end{itemize}

\subsection*{Phase II: Design \& Prototype for Human-AI Collaboration}
Following the theoretical foundation established in Phase I, Phase II combines the assembly of a representative user-story dataset with the implementation of our first working prototype. Incomplete or inconsistent data can lead to unreliable AI outputs, so this phase cleans, normalizes, and streams real-world user stories into the Acceptance Criteria Assistant. Feeding these harmonized inputs into the tool aligns data quality and tool performance, enabling joint iteration and rapid refinement.
\begin{itemize}
    \item \textbf{Objective:} 
    Design human-AI collaboration model for requirements engineering and develop a prototype Acceptance Criteria Assistant for AI‐assisted acceptance criteria generation. 
    \item \textbf{Method:} 
    Conduct internal design reviews to capture prototype requirements and map out human-AI interaction flows. Create wireframes and clickable mockups to visualize the acceptance-criteria workflow. Implement a prototype that integrates multiple LLMs via a unified `generate(model, prompt)` interface. Run pilot tests on the curated user-story dataset, logging AI outputs and user interactions to guide iterative refinements.
    \item \textbf{Expected Outcome:} A functional prototype tool plus an interaction log (user actions, AI outputs, edits) that validates our design choices and surfaces early technical and usability issues.
    \item \textbf{RQ Mapping:} This phase addresses RQ1 by demonstrating AI-assisted acceptance criteria generation in a human-in-the-loop workflow, and RQ2 by identifying integration challenges (prompt design, API orchestration, logging).
\end{itemize}

\subsection*{Phase III: Model Finetuning and Validation}

In this phase, we will use the curated dataset from Phase II to train, fine-tune, and validate AI models. The models assist in various RE activities such as requirements elicitation, classification, and conflict identification. This phase also addresses issues of AI bias, transparency, and practical feasibility before implementing them.
\begin{itemize}
    \item \textbf{Objective:} To devise AI models which are custom-created for RE tasks, with a commitment to ethical accountability, fairness, and transparency. 
    \item \textbf{Method:} Collecting data through advanced AI, such as NLP models (e.g., GPT), for requirements gathering. Use decision-tree or neural network methods in prioritization tasks. Use pre-trained models, Retrieval-Augmented Generation (RAG) architectures\cite{lewis2020retrieval}, and fine-tune the models with domain-specific RE data for better performance in task-specific contexts.
    \item \textbf{Expected Outcome:} Trained AI system, capable of RE tasks such as elicitation, analysis, prioritization, and validation with improved fairness, interpretability, and ethical accountability.
    \item \textbf{RQ Mapping:} This phase addresses RQ3 by focusing on ethical aspects during the model training process, such as reducing bias, enhancing transparency, and ensuring accountability in how AI supports RE tasks.
\end{itemize}

\subsection*{Phase IV: Empirical Experiments \& Evaluation}

Following prototype implementation in Phase II and refinement in Phase III, Phase IV focuses on empirically evaluating our human-AI collaboration approach in real-world requirements engineering settings.
\begin{itemize}
  \item \textbf{Objective:}  
    Empirically assess the effectiveness, trustworthiness, and usability of the tool in real-world RE workflows.
  \item \textbf{Method:}  Conduct stakeholder workshops with 8–10 practitioners using the tool on 3–5 real user stories and rating usability and trust, run controlled case studies on 20 user stories to compare AI-assisted and manual RE by measuring completion time, edit rates, and errors, and perform a longitudinal trial with an engineering team to log usage patterns, feedback submissions, and trust calibration.
  \item \textbf{Expected Outcome:}  A mixed‐methods dataset of quantitative metrics (e.g., edit rates, response latency, SUS and custom survey scores) and qualitative insights (interview notes, usability observations) that reveal the strengths, limitations and areas for refinement of the tool.
  \item \textbf{RQ Mapping:}  
    This phase addresses RQ1 by quantifying AI-assisted acceptance criteria generation gains, RQ2 by surfacing integration and workflow challenges, and RQ3 by evaluating fairness, transparency, and stakeholder acceptance under real-world conditions.
\end{itemize}


\section{Human-AI RE Synergy Model (HARE-SM)}
\label{sec:HARE}



Building on the dataset curation and prototype implementation completed in Phase II and driven by the insights from Phase I, we introduce the \textbf{Human-AI RE Synergy Model (HARE-SM)} to establish a structured and collaborative approach to integrate AI into RE. The model is designed not to replace human expertise but to enhance it through the use of AI for repetitive tasks while ensuring human oversight in critical decision-making.

In Phase I, we have conducted a literature review (currently under peer-review) where we identified four important factors for human-AI collaboration in the context of requirements engineering:

\begin{enumerate}
    \item \textbf{Human-in-the-Loop Validation} -  Avoiding automation bias by ensuring humans make final calls on AI suggestions\cite{fragiadakis2024,mittelstadt2016}.
    \item \textbf{Explainability \& Transparency} - Making AI outputs interpretable so stakeholders understand and trust the reasoning\cite{arrieta2020,ferrari2025formal}.
    \item \textbf{Bias Mitigation}  - Integrating fairness-aware checks to detect and correct skewed outputs\cite{ma2023fairness}.
    \item \textbf{Trust Calibration \& Feedback Loops} - Providing mechanisms for users to give feedback and adjust the AI’s behavior over time\cite{hamza2024,subramonyam2022}.
\end{enumerate}
Guided by these factors and the lessons learned from our Phase II prototype, HARE-SM is based on four key principles that ensure that AI remains a responsible and effective assistant in the requirement process.

\begin{enumerate}
    \item \textbf{Human-in-the-loop validation} - AI assists, but humans make final decisions.
    \item \textbf{Explainability \& Transparency} - AI-generated requirements are made interpretable and traceable.
    \item \textbf{Bias Mitigation Strategies} - The model must enable bias tracing and reveal issues with different LLM techniques and source data.
    \item \textbf{Stakeholder Trust Calibration} - The model should be adjustable for different stakeholders needs and concerns.
\end{enumerate}

\begin{figure}[H]
    \centering
    \includegraphics[width=0.5\textwidth]{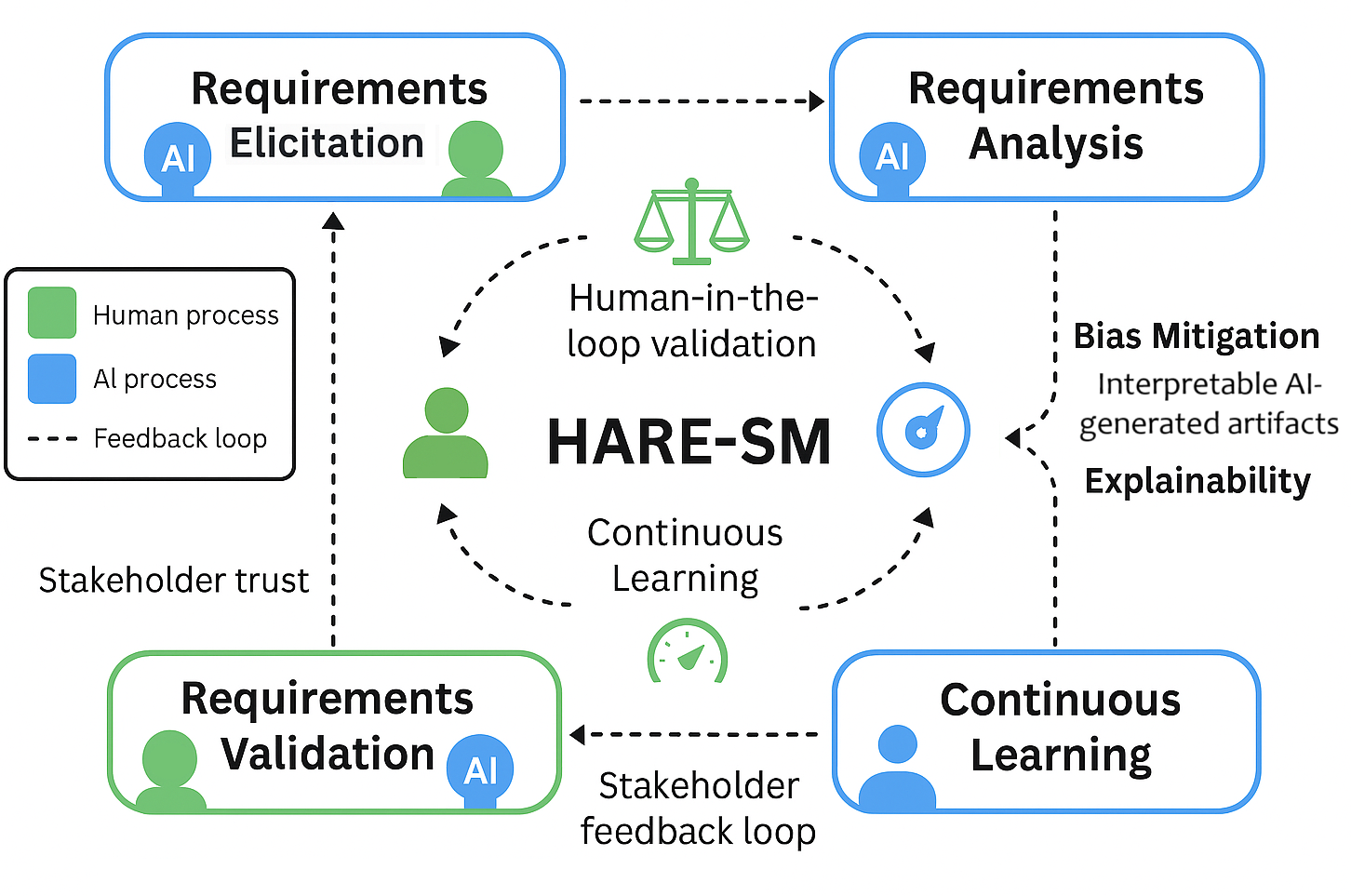}
    \caption{Human-AI RE Synergy Model (HARE-SM) 
    }
    \label{fig:HARE-SM}
\end{figure}

Fig.~\ref{fig:HARE-SM} illustrates the proposed HARE-SM, outlining a structured and collaborative approach for integrating AI into RE. The model offers a balanced distribution of tasks between human expertise and AI automation and addresses concerns such as trust, explainability, and bias mitigation.
HARE-SM consists of the following phases:
\begin{enumerate}
    \item \textbf{Requirements Elicitation:} AI will assist in gathering initial requirements from stakeholders through NLP techniques, identifying ambiguities and inconsistencies for human analysts to review.
    \item \textbf{Requirements Analysis:} LLMs (e.g., Flan-T5, LLaMA-3) and NLP techniques will process the elicited requirements to identify redundancies and potential conflicts, and human experts will validate that these insights match project goals and stakeholder expectations.
    \item\textbf{Requirements Validation:} 
    The validation techniques includes consistency checks and compliance analysis, which will assist humans in refining requirements. 
    Human experts will make sure that AI-generated insights align with stakeholder expectations, mitigating biases and integrating fairness-aware adjustments. This phase is essential for enhancing explainability, maintaining stakeholder trust, and ensuring regulatory compliance.
    \item \textbf{Continuous Learning:} HARE-SM will learn from human feedback and project outcomes to improve its performance over time, fostering an adaptive and evolving collaboration.
\end{enumerate}

\section{HARE-SM Prototype}
\label{sec:approach}

To explore and operationalize the HARE-SM in practice, we developed a functional prototype tool that supports AI-assisted generation and refinement of requirements. In the following we introduce the current features of the prototype and describe how it can be used in RE process.

\subsection{Prototype Features}
It enables users to enter a user story and configure which models to invoke, along with their prompting techniques, parameters, and any contextual metadata. The interface then displays the output of each model side by side, allowing a direct comparison of phrasing, completeness, and specificity, so that the engineer can select or edit the 'best' acceptance criteria. All interactions are recorded in real time, and in later phases these logs will be analyzed to surface common issues (e.g., bias, ambiguity) and generate targeted reports for stakeholders.

The design emphasizes modularity, transparency, and human oversight. Each model’s response is displayed in its own pane, and users can accept, edit, or regenerate outputs using intuitive controls. Session specific data, including response times, selected models, and edit histories, is logged for analysis. The back end captures structured feedback logs in a tabular format (CSV/XLSX) for downstream data-science analysis. This logging mechanism underpins HARE-SM’s core principles by ensuring that every AI suggestion can be reviewed, corrected, and audited.

Functionally, this implementation acts as both a user-facing RE assistant and a data collection engine. As users interact with the tool, feedback logs are generated that capture acceptance decisions, human edits, and preference for particular LLMs. These logs will inform future design iterations and empirical validation of the HARE-SM workflow. The prototype is designed to surface usability patterns, such as preferences for simpler model outputs and challenges in selecting between similar suggestions. These observations will inform the empirical validation planned in Phase IV.

\begin{figure*}[!t]
  \centering
  \includegraphics[width=\linewidth]{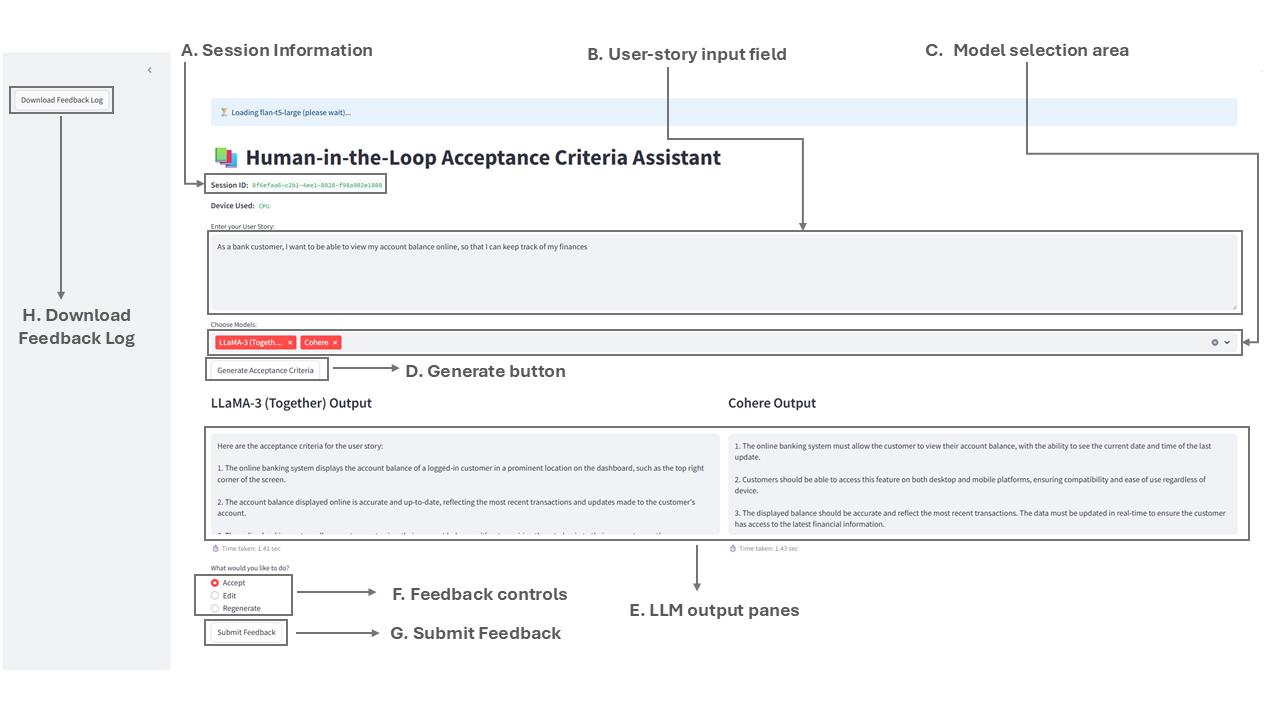}
  \caption{Annotated interface of the Human-in-the-Loop Acceptance Criteria Assistant. 
  (A) Session Information (unique ID and device type). 
  (B) User-story Input Field. 
  (C) Model Selection Area. 
  (D) Generate Button. 
  (E) LLM Output Panels. 
  (F) Feedback Controls. 
  (G) Submit Feedback. 
  (H) Download Feedback Log.}
  \label{fig:ui-annotated}
\end{figure*}

Fig.~\ref{fig:ui-annotated} shows the annotated interface of the prototype tool, detailing each key component of the workflow.


\subsection{System Architecture}
\begin{figure}[ht]
  \centering
  \includegraphics[width=\linewidth]{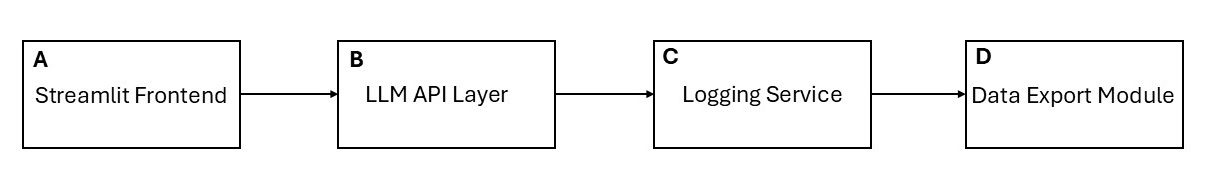}
  \caption{System architecture of the Acceptance Criteria Assistant prototype. 
  (A) Prototype user interface for interactive acceptance criteria generation.
  (B) LLM API layer including tokenization and model servers. 
  (C) Logging service using in-memory session state and optional Excel export for feedback capture. 
  (D) Data export module generating CSV/XLSX logs for analysis.}
  \label{fig:system-arch}
\end{figure}

Fig.~\ref{fig:system-arch} illustrates the end-to-end pipeline behind our prototype. 
The front-end interface collects user stories and model selections. The API layer invokes each LLM and returns raw outputs. A logging service stores outputs, edits, and timestamps in memory via Streamlit’s session state. Finally, the data export module writes structured CSV/XLSX logs for downstream analysis. This modular design enforces HARE-SM’s core principles of human-in-the-loop validation, explainability and transparency, bias mitigation strategies, and trust calibration through feedback loops.

The Logging Service enforces transparency and auditability, the data export module supports explainability and downstream analysis, and the editable front-end enables human-in-the-loop control for trust calibration.

\subsection{Implementation and Model Support}
The Acceptance Criteria Assistant is implemented in Python with Streamlit serving as the front-end framework. It integrates both local and remote large-language models (LLMs) through a unified interface:
\subsubsection*{Local Inference}

\begin{itemize}
    \item We load Google’s Flan-T5 (google/flan-t5-large) using Hugging Face’s transformers library. The model and its tokenizer are instantiated via
    
    tokenizer = AutoTokenizer.from\_pretrained("google/flan-t5-large")

    model     = AutoModelForSeq2SeqLM.from\_pretrained("go\-ogle/flan-t5-large")

    \item Inference runs on CPU or GPU depending on availability. Any other HF-compatible Seq2Seq model can be swapped in by changing the model identifier. 
\end{itemize}
\subsubsection*{API-Backed Models}
\begin{itemize}
    \item Gemini, Cohere, and LLaMA-3 are invoked through their respective REST or SDK APIs (with authentication managed via environment variables), and GPT-3.5 access is routed through OpenRouter’s endpoint to standardize the request format.
\end{itemize}
Each model is wrapped in a common generate(model\_name, prompt) function, and users select which engines to call via a Streamlit multiselect control. Adding new models, whether a different Hugging Face checkpoint or a custom API, requires only registering the model identifier, implementing a small wrapper function, and updating the selection list in the UI. Each model integrated into the prototype presents distinct characteristics. For example, Flan-T5 tends to generalize better but may omit edge cases. Gemini often yields fluent responses but may over-simplify technical details. LLaMA-3 has strong specificity but can be verbose. These characteristics will be evaluated more formally during Phase III. This modular approach keeps the tool adaptable as new LLMs emerge.
\section{Future Work}
\label{sec:Contributions}

Having completed Phases I and II, we now turn to the remaining roadmap activities. Section~\ref{sec:HARE} outlines our plans for Phase III (Model Fine-tuning \& Validation), Phase IV (Empirical Experiments \& Evaluation), and the final synthesis of our results into practical ethical guidelines for AI-augmented RE.
\begin{table*}[t]
  \centering
  \caption{Research Roadmap Status}
  \label{tab:roadmap-status}
  {\footnotesize
   \setlength{\tabcolsep}{4pt}
   \begin{tabularx}{\textwidth}{@{}l X c@{}}
     \toprule
     \textbf{Phase}   & \textbf{Description}                           & \textbf{Status} \\ 
     \midrule
     Phase I: Preliminary Studies  
       & Literature review and gap analysis                               & Completed \\  
     Phase II:  Design \& Prototype for Human-AI Collaboration  
       & Dataset preparation and initial prototype implementation         & Completed \\  
     Phase III: Model Fine-tuning \& Validation  
       & LLM fine-tuning, explainability, and bias mitigation experiments & Planned   \\  
     Phase IV: Empirical Experiments \& Evaluation  
       & Stakeholder workshops, controlled case studies, and field trial   & Planned   \\  
     \bottomrule
   \end{tabularx}
  }
\end{table*}

\begin{itemize}
  \item \textbf{Refinement and Extension of HARE-SM (Phase I\&II):}  
    The initial Human-AI RE Synergy Model has been designed and prototyped; future work will refine its pillars and interaction patterns based on ongoing feedback and lessons learned in the prototype stage.
\end{itemize}

As Table I shows, Phases I and II are complete, while Phases III and IV remain. Our next steps build directly on these planned activities.
\begin{itemize}
  \item \textbf{Phase III – Model Fine-tuning \& Validation:}  
    In Phase III, we will fine-tune our LLMs on the curated dataset, experiment with explainability techniques and bias mitigation strategies, and rigorously validate model outputs against expert criteria.

  \item \textbf{Phase IV – Empirical Experiments \& Evaluation:}  
    Phase IV launches stakeholder workshops, controlled case studies, and a longitudinal field trial to measure productivity, trust, and usability in live RE settings.

  \item \textbf{Ethical and Responsible AI Guidelines (Final Outcome):}  
    We will synthesize our refined model, prototype implementation, and empirical findings into practical guidelines for fairness, transparency, and accountability in AI-augmented requirements engineering.
\end{itemize}

\section{Conclusions}
\label{sec:Conclusion}

The integration of AI in RE brings efficiency, accuracy, and flexibility; however, it also introduces ethical concerns, bias, and transparency challenges in decision-making. This paper proposed the Human-AI RE Synergy Model (HARE-SM), a structured framework for collaborative AI-human workflows in elicitation, analysis, and validation. HARE-SM preserves human oversight by embedding explainability, bias mitigation, and trust calibration mechanisms into every interaction, ensuring that AI amplifies domain expertise without sacrificing accountability. We demonstrated these principles through a prototype Acceptance Criteria Assistant, which captures session data, logs feedback, and provides a testbed for iterative improvement.

Future work will focus on empirical validation of HARE-SM through stakeholder studies and case-based experimentation. Additionally, specific tailoring of the model will be explored for specific industries and projects, especially those with changing, complex requirements. By combining cutting-edge AI with responsible design, HARE-SM paves the way for transparent, trustworthy, and scalable AI-augmented requirements engineering.
\section*{Acknowledgment}
This work has been supported by FAST, the Finnish Software Engineering Doctoral Research Network, funded by the Ministry of Education and Culture, Finland.

\bibliographystyle{IEEEtran}
\bibliography{References} 
\balance
\end{document}